\documentstyle[12pt]{article}
\newcommand{\bm}[1]{\mbox{\boldmath$#1$}}
\newcommand{\text}[1]{\mbox{\scriptsize #1}}
\date{}
\title{Spin-orbital motion and Thomas precession in the classical
and quantum theories}
\author{V.A.~Bordovitsyn~\thanks{E-mail: bord@mail.tomsknet.ru}
and A.N.~Myagkii~\thanks{E-mail: myagkii@mail.ru}\\
\it Tomsk State University, 634050, Tomsk, Russia}

\begin{document}
\maketitle

\begin{abstract}
The motion of a magnetic spin particle in electromagnetic fields is
considered on the basis of general principles of the classical
relativistic theory. Alternative approaches in derivation of the equations
of charge motion and spin precession, the problem of noncollinearity of
the momentum and velocity of a particle with spin, the origin and the
meaning of Thomas precession in dynamics of the spin particle are also
considered.  The correspondence principle in the spin theory is discussed.
\end{abstract}

\section{Introduction}

\hspace*{\parindent}
In the initial form the classical spin theory was constructed by
Ya.I.~Frenkel~\cite{1} almost at the same time when the spin was discovered.
The main theses of quantum spin theory were formulated by W.~Pauli.
Later with appearance of the Dirac relativistic equation the quantum
spin theory was consecutively developed.

For a long time the classical and quantum spin theories developed
independently from each other. It seemed that there was no connection between
them. A large number of authors, including the authors of well-known
books~\cite{2,3,4}, assumed that there was no the classical spin theory in
general (see also~\cite{5}).

Later the Bargmann-Michel-Telegdi (BMT) classical spin equation~\cite{6} in
accordance with the Frenkel spin equation (see~\cite{7})
was obtained on the basis of the relativistic quasiclassical
spin theory~\cite[9-11]{8} and the Heisenberg equations of operators
motion in the relativistic quantum mechanics (see~\cite{12}). After that
the situation has changed.  BMT equation successfully confirmed the
results of experiments on precision determination of the anomalous
magnetic moment of an electron~\cite{13,14}, muon~\cite{15} and also the
masses of a number of heavy elementary particles~\cite{16}. As it turned
out well to explain the experimentally observed effect of radiative
self-polarization of electrons~\cite{17,18} within the frameworks of the
quasiclassical spin theory as well as the Dirac quantum theory. Moreover,
it was found that even purely quantum effect such as spin-flip radiation
has classical description~\cite{18.1}.

Nevertheless there is no united point of view about the fundamental
principles of construction of the classical spin theory in the modern
scientific and educational literature. There is a great variety of the
classical methods of spin description, which are not confirmed with each
other in many respects (see, for example, \cite{19}). In this work the
attempt in constructing of a correct classical spin theory was
made on the basis of the general principles of classical electrodynamics.
Of course, the classical and the quantum spin theories have to be correlated
at $\hbar\to 0$.

A discussion of this problem is actual because many problems of
the spin properties of elementary particles (noncollinearity of the particle
momentum and velocity, the origin of the Thomas precession and
emission of spin particles etc.) can be apparently resolved only by means
of the classical theory. The correctly formulated classical spin theory allows
also to understand more profound some of purely quantum properties of spin
particles as phenomena of creation of pairs and Zitterbewegung, the
problem of maximal acceleration et al.

\section{Equations of motion of a spin particle in the classical
electrodynamics}

\hspace*{\parindent}
Here we obtain the equations of motion of a classical spin particle by means
of the Lagrange method and the energy-momentum tensor formalism.

\subsection{Lagrangian formulation of the problem}

\hspace*{\parindent}
We introduce the Lagrange function of a charged particle with intrinsic
magnetic moment in the additive form
$$L=L_e+L_s,$$
where
$$L_e=\frac{1}{2}\;m_0v_\rho v^\rho+\frac{e}{c}\;v_\rho A^\rho$$
is the well-known Lagrange function of the relativistic charged particle (see,
for example, \cite{20}),
$$L_s=\frac{\mu}{2}\;H_{\alpha\beta}\Pi^{\alpha\beta}$$
is the spin addition taking into consideration the interaction between the
intrinsic magnetic moment of the particle and an external electromagnetic
field.

In the adduced above relations the following notations were taken:
$v^\rho=(v^0,{\bm v})=c\gamma(1,{\bm\beta})$ is the four-dimensional velocity
of a particle, $\gamma=1/\sqrt{1-\beta^2}$ is the Lorentz-factor,
$A^\rho=(\varphi,{\bm A})$ are the potentials of an external field,
$H^{\mu\nu}=\partial^\mu A^\nu-\partial^\nu A^\mu$ is the electromagnetic
field tensor. We take the metric $g^{\mu\nu}=(-1,1,1,1)$ and
therefore $v_\rho v^\rho=-c^2$, $H^{10}=-E_x$, $H^{12}=H_z$ etc. Spin
tensor $\Pi^{\alpha\beta}$ is separated out the tensor of the intrinsic
magnetic moments $\mu^{\alpha\beta}$ (the Frenkel tensor) and therefore it
is dimensionless $$\mu^{\alpha\beta}=\mu\Pi^{\alpha\beta}, \quad
\mu=g\mu_0s.$$

Here $\mu_0=e\hbar/2m_0c$ is the Bohr magneton, $s$ is the spin number, which
we put equal to $1/2$. It is supposed that $g$-factor contains
the anomalous part. As known $g=2(1+\alpha/2\pi)$ for an electron.

The tensor $\Pi^{\alpha\beta}$ is space-like and satisfies the Frenkel
condition~\cite{1}
\begin{equation}
v_\alpha\Pi^{\alpha\beta}=0.
\label{o}
\end{equation}

The more acceptable functional dependence in $L$ is given by the formally
equivalent expression
\begin{equation}
L=\frac{1}{2}\;m_0\left(1-\frac{\mu}{2m_0c^2}\;H_{\alpha\beta}
\Pi^{\alpha\beta}\right)v_\rho v^\rho+\frac{e}{c}\;v_\rho A^\rho+
\frac{\mu}{4}\;H_{\alpha\beta}\Pi^{\alpha\beta}.
\label{L}
\end{equation}
The first term contains a factor
\begin{equation}
m=m_0\left(1-\frac{\mu}{2m_0c^2}\;H_{\alpha\beta}\Pi^{\alpha\beta}\right),
\label{m}
\end{equation}
which can be presented as the mass of a particle with
spin. The spin mass (\ref{m}) was introduced in the works of many authors
(see \cite[22-31]{7,19} et al.) by various means.

Substituting (\ref{L}) into the Euler-Lagrange relativistic equations
we obtain
\begin{equation}
\frac{d}{d\tau}\left(mv^\rho\right)=\frac{e}{c}\;H^{\rho\lambda}v_\lambda+
\frac{\mu}{2}\;\Pi_{\alpha\beta}\partial^\rho H^{\alpha\beta},
\label{vd1}
\end{equation}
where $\tau$ is the proper time. The equation (\ref{vd1}) can be also presented in the
form
\begin{equation}
m\;\frac{dv^\rho}{d\tau}=\frac{e}{c}\;H^{\rho\lambda}v_\lambda+
\frac{\mu}{2}\;\Pi_{\alpha\beta}\partial^\rho H^{\alpha\beta}-
\frac{dm}{d\tau}\;v^\rho.
\label{vd/}
\end{equation}

The calculation of $dm/d\tau$ can be carried out only after formulation of spin
equations of motion (see further) and then we can obtain with an accuracy
to terms of orders $\hbar^2$, $\hbar(g-2)$
$$\frac{dm}{d\tau}=-\frac{\mu}{2c^2}\;v_\lambda\Pi_{\alpha\beta}
\partial^\lambda H^{\alpha\beta}.$$

In this case equation (\ref{vd/}) really takes the space-like form (see \cite{7})
\begin{equation}
m\;\frac{dv^\rho}{d\tau}=\frac{e}{c}\;H^{\rho\lambda}v_\lambda+
\frac{\mu}{2}\;\Pi_{\alpha\beta}
\stackrel{\rightarrow}{\partial}\!\!{}^\rho H^{\alpha\beta},
\label{vd2}
\end{equation}
where
$$\stackrel{\rightarrow}{\partial}\!\!{}^\rho=\partial^\rho+\frac{1}{c^2}\;v^\rho v_\lambda
\partial^\lambda$$
is the space-like derivative.

The equation (\ref{vd1}) was obtained by Frenkel~\cite{1} in the other form by means
of the Lagrange variational multipliers.

\subsection{Energy-momentum tensor formalism}

\hspace*{\parindent}
The density of the energy-momentum tensor of a spin particle
is represented as (radiation field is neglected)
$${\cal P}^{\alpha\beta}={\cal P}^{\alpha\beta}_{e}+
{\tilde{\cal P}}^{\alpha\beta},$$
where
\begin{equation}
{\cal P}^{\alpha\beta}_{e}=c\int^{\infty}_{-\infty}
d{\tau}^{\prime}mv^{\alpha}(\tau^{\prime})v^{\beta}(\tau^{\prime})
\delta\left(R^{\rho}-r^{\rho}(\tau^{\prime})\right)
\label{pe}
\end{equation}
is the energy-momentum tensor density of the charge,
$\tau^{\prime}$ is the invariant parameter
which after integration with $\delta$-function
$$\delta\left(cT-ct(\tau^{\prime})\right)=\frac{\delta(\tau-\tau^{\prime})}
{c{\left(\frac{dt}{d\tau^{\prime}}\right)}_{T=t}}=\frac{1}{c\gamma}
\delta(\tau-\tau^{\prime})$$
gives the proper time $\tau$:
\begin{equation}
{\cal P}^{\alpha\beta}_{e}=mv^{\alpha}(\tau)u^{\beta}(\tau)
\delta\left(\bm R-\bm r(\tau)\right),
\label{_pe}
\end{equation}
where $u^{\beta}(\tau)=(c,\bm u)$, $\bm u$ is ordinary velocity of the
particle.

The formula (\ref{pe}) is written under assumption that a force acting
on the particle is defined by the relation $F^{\alpha}=mw^{\alpha}$.
In the general case
$F^{\alpha}=dP^{\alpha}\!\left/d\tau\right.$, where $P^{\alpha}$ is
the momentum of the particle. Then instead of (\ref{pe}) and
(\ref{_pe}) we have \begin{equation} {\cal
P}^{\alpha\beta}_{e}=c\int^{\infty}_{-\infty}d\tau^{\prime}
P^{\alpha}(\tau^{\prime})v^{\beta}(\tau^{\prime})\delta\left(R^{\rho}-
r^{\rho}(\tau^{\prime})\right)
\label{pe/}
\end{equation}
and
\begin{equation}
{\cal P}^{\alpha\beta}_{e}=P^{\alpha}(\tau)u^{\beta}\delta\left(R^{\rho}-
r^{\rho}\right).
\label{_pe/}
\end{equation}

The energy-momentum tensor of electromagnetic field is specified,
in the usual fashion, by expression
$${\tilde{\cal P}}^{\alpha\beta}=-\frac{1}{4\pi}\left(H^{\alpha\rho}H_{\rho}{}^{\beta}+
\frac{1}{4}g^{\alpha\beta}H_{\rho\gamma}H^{\rho\gamma}\right).$$

In the differential form the conservation energy-momentum law has the form
$$D_{\beta}{\cal P}^{\alpha\beta}=0,$$
where $D_{\beta}=\partial\!\left/\partial R^{\beta}\right.$.
Using the ordinary manner (see, e.g.,~\cite{31})
one can integrate the expression over the four-volume $d\Upsilon$.
Then omitting details of calculations
\footnote{Here we take into account that total current
density is made up the current density of the charge $j^{\alpha}_e$ and
the current density of the magnetic moment $j^{\alpha}_s$, that is
$$j^{\alpha}=j^{\alpha}_{e}+j^{\alpha}_{s},$$
where
$$j^{\alpha}_e=ec\int d\tau^{\prime}
v^{\alpha}(\tau^{\prime})\delta\left(R^{\rho}-r^{\rho}(\tau^{\prime})
\right),\quad
j^{\alpha}_s=\mu c^2\int d\tau^{\prime}
\Pi^{\alpha\lambda}(\tau^{\prime})D_{\lambda}
\delta\left(R^{\rho}-r^{\rho}(\tau^{\prime})\right).$$
} we obtain
\begin{equation}
\frac{d}{d\tau}mv^{\alpha}=\frac{e}{c}H^{\alpha\beta}v_{\beta}+\frac{\mu}{2}
\Pi_{\rho\lambda}\partial^{\alpha}H^{\rho\lambda}
\label{vd}
\end{equation}
in the case (\ref{_pe}) and
\begin{equation}
\frac{dP^{\alpha}}{d\tau}=\frac{e}{c}H^{\alpha\beta}v_{\beta}+\frac{\mu}{2}
\Pi_{\rho\lambda}\partial^{\alpha}H^{\rho\lambda}
\label{pd}
\end{equation}
in the case (\ref{_pe/}).

It is easy to see that the first equation (\ref{vd}) coincides with
the equation obtained by the Langrangian method. The second equation
(\ref{pd}), generally speaking, differs from the above equation
because the momentum $P^{\alpha}$ is not determined.
In the following we shall use the more general second
case.  As we shall see later, the problem of agreement between
equations (\ref{vd}) and (\ref{pd}) is not trivial and will be
discussed below.

\section{Spin equations and the problem of noncollinearity of
         momentum and velocity}

\hspace*{\parindent}
Let us start by considering the successive derivation of the spin equations
in the classical relativistic mechanics.

\subsection{Formalism of angular momentum tensor}

\hspace*{\parindent}
We introduce the density tensor of total angular momentum of a
particle in the electromagnetic field
$$M^{\alpha\lambda\beta}=R^{[\alpha}{\cal P}^{\lambda\beta]}=R^{\alpha}{\cal P}^{\lambda\beta}-
R^{\beta}{\cal P}^{\lambda\alpha}.$$

The conservation law of angular momentum in differential form is
$$D_{\lambda}M^{\alpha\lambda\beta}=0.$$

If we then act as in section 2.2. one obtains the following equation of
motion of the angular momentum $M^{\alpha\beta}$
$$\frac{dM^{\alpha\beta}}{d\tau}=\mu H^{[\alpha\rho}\Pi_{\rho}{}^{\beta]}+
r^{[\alpha}\left(\frac{e}{c}H^{\beta]\rho}v_{\rho}+\frac{\mu}{2}H_{\rho\lambda}
\partial^{\beta]}\Pi^{\rho\lambda}\right),$$
or, taking into account the momentum form of the equation of motion
(\ref{pd}),
\begin{equation}
\frac{dM^{\alpha\beta}}{d\tau}=\frac{d}{d\tau}\left(r^{[\alpha}P^{\beta]}
\right)-v^{[\alpha}P^{\beta]}+\mu H^{[\alpha\rho}\Pi_{\rho}{}^{\beta]}.
\label{md}
\end{equation}

Using the definition of orbital momentum
$$L^{\alpha\beta}=r^{\alpha}P^{\beta}-r^{\beta}P^{\alpha}$$
we can represent the equation (\ref{md}) as
$$\frac{dM^{\alpha\beta}}{d\tau}=\frac{d}{d\tau}\left(L^{\alpha\beta}+
S^{\alpha\beta}\right).$$

It follows that $S^{\alpha\beta}$ is the spin moment of the
particle.  By separating from $S^{\alpha\beta}$ the factor $\hbar/2$
we come to the dimensionless spin equation
\begin{equation}
\frac{\hbar}{2}\frac{d\,\Pi^{\alpha\beta}}{d\tau}=-v^{[\alpha}P^{\beta]}+
\mu H^{[\alpha\rho}\Pi_{\rho}{}^{\beta]}.
\label{pid}
\end{equation}

It is also evident that
$\Pi_{\alpha\beta}\left(d\Pi^{\alpha\beta}\!\!\left/d\tau\right.\right)=0$,
i.e.  $\Pi_{\alpha\beta}\Pi^{\alpha\beta}=\rm const$ is well known
spin invariant.  We emphasize that equation (\ref{pid}) was derived
under assumption that the Frenkel condition (\ref{o}) is valid.

\subsection{Spin equations in the Frenkel formalism}

\hspace*{\parindent}
If, as usual, the momentum of a particle is defined by relation
$P^{\alpha}=mv^{\alpha}$, i.e. momentum and velocity are collinear, the spin
equation (\ref{pid}) can be simplified to
$$\frac{\hbar}{2}\frac{d\,\Pi^{\alpha\beta}}{d\tau}=\mu H^{[\alpha\rho}
\Pi_{\rho}{}^{\beta]}.$$

This equation differs essentially from experimentally verified BMT equation
since formally it does not contain in the pure state the terms with
anomalous magnetic moment
$\mu_a=\mu-\mu_0=\mu_0(g-2)/2$. Matters can be improved by replacement
\begin{equation}
P^{\alpha}=mv^{\alpha}+\stackrel{\rightarrow}{Z}\!{}^{\alpha}.
\label{p}
\end{equation}
Additional term $\stackrel{\rightarrow}{Z}\!{}^{\alpha}$ must be concerned
with anomalous magnetic moment in a way. In general, it leads
to noncollinearity of the momentum and the velocity of the spin
particle.  Taken along, this fact is not unexpected. It is reasonable
that the spin particle moving along curved trajectory executes a
peculiar trembling motion (Zitterbewegung).

Notice that the relation of the type (\ref{p}) was first
adduced in the original Frenkel theory \cite{1}. Thus we can say
that the first definition of spin mass $m$ was made by
Ya.I.Frenkel ($m$ is the Langrange uncertain multiplier
$\lambda$ in the Frenkel theory).

Next we shall assume that the additional term
$\stackrel{\rightarrow}{Z}\!{}^{\alpha}$ is space-like vector which
satisfies condition
\begin{equation}
v_{\alpha}\stackrel{\rightarrow}{Z}\!{}^{\alpha}=0.
\label{vz}
\end{equation}

It follows that in the system of rest the particle energy is equal
to $mc^2$ and contains the potential energy of the intrinsic magnetic
moment of the particle (see definition $m$ in (\ref{m})). According
to (\ref{p}) and (\ref{vz}) we obtain
\begin{equation}
v_{\alpha}P^{\alpha}=-mc^2.
\label{vp}
\end{equation}

By substituting (\ref{p}) into the spin equation (\ref{pid}) we have
\begin{equation}
\frac{\hbar}{2}\frac{d\,\Pi^{\alpha\beta}}{d\tau}=-v^{[\alpha}
\stackrel{\rightarrow}{Z}\!{}^{\beta]}+\mu H^{[\alpha\rho}
\Pi_{\rho}{}^{\beta]}.
\label{pid/}
\end{equation}

It is interesting that the spin mass drops out of the equation. To
determine $\stackrel{\rightarrow}{Z}\!{}^{\alpha}$ equation (\ref{pid/})
should be multiplied by $v_{\beta}$. Then we find \begin{equation}
\stackrel{\rightarrow}{Z}\!{}^{\alpha}=\frac{\hbar}{2c^2}\Pi^{\alpha\beta}
w_{\beta}-\frac{\mu}{c^2}\Pi^{\alpha}{}_{\rho}H^{\rho\beta}v_{\beta}.
\label{pid/1}
\end{equation}

Rewriting $\stackrel{\rightarrow}{Z}\!{}^{\alpha}$ with regard for the
equation of motion (\ref{vd2}) and the Frenkel condition (\ref{o}) we
obtain \begin{equation}
\stackrel{\rightarrow}{Z}\!{}^{\alpha}=\Pi^{\alpha\beta}\left(
\frac{\mu\hbar}{4mc^2}\Pi_{\rho\lambda}\stackrel{\rightarrow}{\partial}_{\beta}H^{\rho\lambda}-
\frac{\mu_a}{c^2}H_{\beta\rho}v^{\rho}\right).
\label{_z}
\end{equation}

Thus, here we can see not only the expected anomalous magnetic
moment $\mu_a$ but also a gradient in field term which is
proportional to $\hbar$.
Since $\stackrel{\rightarrow}{Z}\!{}^{\alpha}\sim\hbar^2,\;
\hbar(g-2)$ the spin mass in (\ref{_z}) is replaced by the
rest mass $m_0$.

It is significant from definition of
$\stackrel{\rightarrow}{Z}\!{}^{\alpha}$
(\ref{_z}) it follows that a free particle
does not undergo the spin perturbation of trajectory. If we
assume that $g=2$ (anomalous magnetic moment is absent)
$\stackrel{\rightarrow}{Z}\!{}^{\alpha}$ will coincide with the expression
found by Frenkel\footnote{In notation of Frenkel we have
(see formula (22a) in Ref.~\cite{1})
$$\stackrel{\rightarrow}{Z}\!{}^{\alpha}=-\mu^{\alpha\beta}a_{\beta},\quad
a^{\beta}=-\frac{1}{2e_0c}\mu_{\rho\lambda}\partial^{\beta}H^{\rho\lambda}.$$
} in a different way
$$\stackrel{\rightarrow}{Z}\!{}^{\alpha}\left|_{g=2}\right.=\frac{1}{4}
\frac{\mu_0\hbar}{m_0c^2}\Pi^{\alpha\beta}\Pi_{\rho\lambda}\stackrel{\rightarrow}{\partial}_{\beta}
H^{\rho\lambda}.$$
For arbitrary values of $g$ the expressions
$\stackrel{\rightarrow}{Z}\!{}^{\alpha}$ in (\ref{pid/1}) and (\ref{_z})
are associated with definition of the momentum in the Corben
theory~\cite{23}.

By substituting $\stackrel{\rightarrow}{Z}\!{}^{\alpha}$ (\ref{_z}) into
equation (\ref{pid/}) we obtain the spin equation
\begin{equation}
\frac{d\,\Pi^{\alpha\beta}}{d\tau}=\frac{eg}{2m_0c}H^{[\alpha\rho}
\Pi_{\rho}{}^{\beta]}+v^{[\alpha}\Pi^{\beta]\rho}\left(
\frac{g-2}{2}\frac{e}{m_0c^3}H_{\rho\lambda}v^{\lambda}-\frac{\mu}{2m_0c^2}
\Pi_{\eta\lambda}\stackrel{\rightarrow}{\partial}_{\rho}H^{\eta\lambda}\right).
\label{_z/}
\end{equation}

This equation is written with an accuracy to terms of orders $\hbar$,
$g-2$.  The spin equation (\ref{_z/}) was first obtained by
H.Corben~\cite{23} (see also \cite{20,33}).  In particular case of
uniform fields this equation transforms to the tensor form of the BMT spin
equation (see~\cite{7,34}).

Now we can refine the expression $dm/d\tau$ in the equation (\ref{vd/}).
According to (\ref{vp}) and (\ref{pd}) we
have~\footnote{The exact expression $dm/d\tau$ has the form
(see also~\cite{35})
$$\frac{dm}{d\tau}=\frac{\mu}{c^4}w_{\alpha}\Pi^{\alpha\rho}H_{\rho\beta}
v^{\beta}-\frac{\mu}{2c^2}\Pi_{\rho\lambda}v_{\alpha}\partial^{\alpha}
H^{\rho\lambda}.$$
However, it can be shown that the first term has just an accuracy of
$\hbar^2$, $\hbar(g-2)$.
}
$$\frac{dm}{d\tau}=-\frac{1}{c^2}\left(w_{\alpha}P^{\alpha}+v_{\alpha}
\frac{dP^{\alpha}}{d\tau}\right)\approx-\frac{\mu}{2c^2}\Pi_{\rho\lambda}
v_{\alpha}\partial^{\alpha}H^{\rho\lambda}.$$

This expression is written with an accuracy to $\hbar^2$,
$\hbar(g-2)$ since we assume that the equation of charge motion has an
accuracy of the first order in $\hbar$.

Hence, if we substitute $dm/d\tau$ in the equation of charge motion
(\ref{vd/}) it will taken the form of the equation (\ref{vd2}) considered in
subsection 2.1.

\subsection{Spin equation in the Shirokov formalism}

\hspace*{\parindent}
Instead of the Frenkel condition (\ref{o}) many authors (see, for example,
in Ref.~\cite{19,36}) use the auxiliary condition introduced by Yu.M.Shirokov~\cite{37}
\footnote{It will be noted that the condition (\ref{m}) was first
formulated by Yu.M.Shirokov for spin four-vector. As it is known
(see, for example,~\cite{32}) the spin four-vector $S^{\mu}$ is connected
with tensor $\Pi^{\alpha\beta}$ by relation
$$S^{\mu}=\frac{1}{2m_0c}\varepsilon^{\mu\nu\alpha\beta}\Pi_{\alpha\beta}
P_{\nu},\quad
\Pi^{\alpha\beta}=\frac{1}{m_0c}\varepsilon^{\alpha\beta\rho\sigma}
S_{\rho}P_{\sigma}.$$
It follows that in the Shirokov formalism condition (\ref{m}) have the form
$P_{\mu}S^{\mu}=0$.
Sometimes the condition (\ref{m}) is called as Nakano condition~\cite{38}
(see also~\cite{39}).}
\begin{equation}
P_{\alpha}\Pi^{\alpha\beta}=0.
\label{sh}
\end{equation}

In this case the intrinsic angular momentum
is specified with respect to the inertial center of the spin
particle in the covariant form.  The classical Zitterbewegung is
a result of the fact that coordinates of the inertial center do not
coincide with coordinates of the particle.

On the lines of Shirokov formalism the expansion of the momentum
(\ref{p}) can be rewritten in the form
\begin{equation}
P^{\alpha}=mv^{\alpha}+Z^{\alpha},
\label{p/}
\end{equation}
where vector $Z^{\alpha}$ satisfies the condition~\cite{40}
\begin{equation}
P_{\alpha}Z^{\alpha}=0.
\label{pz}
\end{equation}

By analogy with preceding case we obtain instead of (\ref{pid/1})
the following expression
\begin{equation}
Z^{\alpha}=\frac{m}{P_{\alpha}P^{\alpha}}\left(-\frac{\hbar}{2}
\Pi^{\alpha\beta}\frac{dP_{\beta}}{d\tau}+\mu\Pi^{\alpha}{}_{\rho}
H^{\rho\beta}v_{\beta}\right).
\label{z}
\end{equation}

With due regard for smallness of square
$Z_{\alpha}Z^{\alpha}\sim\hbar^2$ we have
\begin{equation}
P_{\alpha}P^{\alpha}\approx-m^2c^2.
\label{pp}
\end{equation}

It should be noted that this relation remains valid also within the
framework of Frenkel formalism.

Using the equation of motion (\ref{pd}), conditions (\ref{pz}) and
relations(\ref{pp}) we come to the following expression
\begin{equation}
Z^{\alpha}=\Pi^{\alpha\beta}\left(\frac{\mu\hbar}{4m_0c^2}\Pi_{\rho\lambda}
\partial_{\beta}H^{\rho\lambda}-\frac{\mu_a}{c^2}H_{\beta\rho}v^{\rho}\right).
\label{z1}
\end{equation}
for the linear approximation in field.

The expression (\ref{z1}) differs from
$\stackrel{\rightarrow}{Z}\!{}^{\alpha}$ in (\ref{_z}) only by
derivative (there is $\partial^{\alpha}$ instead of
$\stackrel{\rightarrow}{\partial}\!\!{}^{\alpha}$).
However, it is unimportant what
derivative presents in the spin equations in the Frenkel formalism,
$\stackrel{\rightarrow}{\partial}\!\!{}^{\alpha}$ and $\partial^{\alpha}$, as
$$\Pi_{\alpha\beta}\partial^{\beta}=\Pi_{\alpha\beta}
\stackrel{\rightarrow}{\partial}\!\!{}^{\alpha}.$$

It is very interesting that in the Shirokov formalism the spin equation
(\ref{pid}) has the same form as in the Frenkel formalism.
Therefore, in the Shirokov and Frenkel formulations the spin equations
coincide with the equation (\ref{_z/}) up to terms linear in the external
field.  As far as the charge equation of motion are concerned they
coincide with equations (\ref{vd}) and (\ref{pd}) in the linear
approximations in powers of $\hbar$ and in field (see also in
Ref.~\cite{35}, p.51).

\section{Thomas precession as a kinematical spin-orbit effect}

\hspace*{\parindent}
It is best to perform the analysis of the spin equations of motion by
use of spin vector $\bm\zeta$ given in the rest frame. This vector is
connected with components of the tensor $\Pi^{\alpha\beta}$ by means
of Lorentz transformation
$$\bm\zeta=\frac{1}{\gamma}{\bm\Pi}+\frac{\gamma}{\gamma+1}\bm\beta
(\bm\beta\bm\Pi)$$
and the corresponding spin equation has the form
\begin{eqnarray}
\frac{d\bm\zeta}{d\tau}&=&\frac{eg}{2m_0c}[\bm\zeta\bm H]-
\frac{e(g-2)}{2m_0c}\frac{\gamma^2}{\gamma+1}
[\bm\zeta[\bm\beta[\bm\beta\bm H]]]-\frac{e}{m_0c}\gamma\left(\frac{g}{2}-
\frac{\gamma}{\gamma+1}\right)[\bm\zeta[\bm\beta\bm E]]+ \nonumber \\
&+&\frac{ge\hbar}{4m^2_0c^2}\frac{\gamma^2}{\gamma+1}[\bm\zeta[\bm\beta\bm\nabla]]
\left\{(\bm\zeta\bm H)+(\bm\beta[\bm\zeta\bm E])-\frac{\gamma}{\gamma+1}
(\bm\beta\bm\zeta)(\bm\beta\bm H)\right\}.
\label{a}
\end{eqnarray}

This equation is known in literature (see, for instance, \cite{41}) for
particular case of homogeneous fields. In this case the equation of spin
precession can be presented in the form
\begin{equation}
\frac{d{\bm\zeta}}{dt}=[{\bm\Omega\bm\zeta}].
\label{a1}
\end{equation}

In some papers (see, for instance, \cite{42,43})) it was shown that the
precession frequency ${\bm\Omega}$ consists of two fundamentally different
parts
$${\bm\Omega}={\bm\Omega}_{\rm L}+{\bm\Omega}_{\rm Th},$$
where
\begin{equation}
{\bm\Omega}_{\rm L}=-\frac{eg}{2m_0c}\left(\frac{1}{\gamma}\;{\bm H}-
[{\bm\beta}{\bm E}]-\frac{\gamma}{\gamma+1}[{\bm\beta}[{\bm\beta}{\bm H}]]
\right)
\label{OL}
\end{equation}
is the Larmor frequency and
\begin{equation}
{\bm\Omega}_{\rm Th}=-\frac{e}{m_0c}\;\frac{\gamma}{\gamma+1}
\left([{\bm\beta}{\bm E}]+[{\bm\beta}[{\bm\beta}{\bm H}]]\right)
\label{Th}
\end{equation}
is the Thomas precession frequency which, as known, has the kinematical origin
$${\bm\Omega}_{\rm Th}=-\frac{1}{c}\;\frac{\gamma^2}{\gamma+1}\;
[{\bm\beta}{\bm a}],$$
where ${\bm a}$ is the particle acceleration.

Transforming in equations (\ref{OL}) and (\ref{Th}) the electromagnetic
fields to the rest frame of a particle we obtain the simple result
$${\bm\Omega}_{\rm L}=-\frac{eg}{2m_0c}\;\frac{1}{\gamma}\;{\bm H}_0, \quad
{\bm\Omega}_{\rm Th}=-\frac{e}{m_0c}\;\frac{1}{\gamma+1}\;
[{\bm\beta}{\bm E}_0].$$

The Thomas precession~\cite{44} is considered in a large number
papers (see~\cite[43-48]{24} et al.). Here we want to show how the separation of
spin equations into dynamical and kinematical parts can be carried out via
covariant form of spin equation (\ref{_z/}). For simplicity, as above, we limit
ourselves by the case of homogeneous fields (BMT approximation).

The equation (\ref{_z/}) is considerably simplified via introduction of a
certain effective
field with physically distinguished kinematical part which meets the Thomas
precession
\begin{equation}
H^{\alpha\beta}_{\rm eff}=\stackrel{\rightarrow}{H}{}^{\alpha\beta}+
\frac{2m_0}{egc}\;v^{[\alpha}w^{\beta]},
\label{a2}
\end{equation}
where
$$w^\beta=\frac{e}{m_0c}\;H^{\beta\rho}v_\rho,$$
$\stackrel{\rightarrow}{H}{}^{\alpha\beta}$ is the space-like part of
an electromagnetic field
$$\stackrel{\rightarrow}{H}{}^{\alpha\beta}=H^{\alpha\beta}+\frac{1}{c^2}\;
v^{[\alpha}v_\rho H^{\rho\beta]}, \quad
v_\alpha\stackrel{\rightarrow}{H}{}^{\alpha\beta}=0.$$

Then the spin equation (\ref{_z/}) takes the simple form
$$\frac{d\,\Pi^{\mu\nu}}{d\tau}=\frac{eg}{2m_0c}H^{[\mu\rho}_{\rm eff}
\Pi_\rho{}^{\nu]}.$$

The equation (\ref{a}) for vector ${\bm\zeta}$ is also considerably simplified
$$\frac{d{\bm\zeta}}{d\tau}=\frac{eg}{2m_0c}\left[{\bm\zeta}\left(
{\bm H}_{\rm eff}-\frac{\gamma}{\gamma+1}[{\bm\beta}{\bm E}_{\rm eff}]
\right)\right].$$

Separating here the kinematical part of fields according to (\ref{a2})
$${\bm H}_{\rm eff}=\stackrel{\rightarrow}{\bm H}+\frac{2}{g}\;\gamma^2
[{\bm\beta}({\bm E}+[{\bm\beta}{\bm H}])],$$
$${\bm E}_{\rm eff}=\stackrel{\rightarrow}{\bm E}+\frac{2}{g}\;\gamma^2
({\bm E}+[{\bm\beta}{\bm H}]-{\bm\beta}({\bm\beta}{\bm E})),$$
where $\stackrel{\rightarrow}{\bm H}$ and $\stackrel{\rightarrow}{\bm E}$
are components of $\stackrel{\rightarrow}{H}{}^{\alpha\beta}$,
we obtain equation (\ref{a1}) with separated frequencies.

Thus one can compare the kinematical field
$$H^{\alpha\beta}_{\rm Th}=\frac{2m_0}{egc}\;v^{[\alpha}w^{\beta]}$$
with Thomas precession.

One can also introduce the kinematical field dual to $H^{\alpha\beta}_{Th}$
$$E^{\alpha\beta}_{\rm Th}=\frac{1}{2}\;\varepsilon^{\alpha\beta\rho\sigma}
H^{\rm Th}_{\rho\sigma}=
\frac{2m_0}{egc}\;\varepsilon^{\alpha\beta\rho\sigma}
w_\rho v_\sigma.$$

It is interesting that $E^{\alpha\beta}_{\rm Th}$ is space-like
tensor:
$$v_\alpha E^{\alpha\beta}_{\rm Th}=0,$$
and
$$E^{\alpha\beta}_{\rm Th}H_{\alpha\beta}^{\rm Th}=0.$$

Another invariant of kinematical field is proportional to the square of
acceleration
$$H^{\alpha\beta}_{\rm Th}H_{\alpha\beta}^{\rm Th}=-\frac{8m_0^2}{e^2g^2}\;
w_\mu w^\mu.$$

We note also that
\begin{equation}
\frac{1}{2}\;H_{\mu\nu}\Pi^{\mu\nu}=
\frac{1}{2}\;\stackrel{\rightarrow}{H}_{\mu\nu}\Pi^{\mu\nu}=
\frac{1}{2}\;H^{\mu\nu}_{\rm eff}\Pi_{\mu\nu}=({\bm\zeta}{\bm H}_0).
\label{i}
\end{equation}

Hence it follows that the Thomas precession gives no addition to the
effective mass (\ref{m}).

Thus we satisfy ourselves that the Thomas precession is a purely kinematical
spin-orbit effect.

\section{Correspondence principle in the spin theory}

\hspace*{\parindent}
In the quantum theory of the Dirac particles the spin description is given
via the Poincare-invariant spin operators~\cite{32}.
Thereat the operator equations
of motion are obtained from the Heisenberg proper time equations with the
Dirac covariant Hamiltonian~\cite{12}. To found a correspondence between the
classical and quantum equations of motion
it is necessary to
separate out the terms responsible for quantum Zitterbewegung. We remind
that this phenomenon unlike the classical Zitterbewegung (see above) is
accounted for interference of charge conjugate states (see, for example,
\cite{48.0, 48} et al.).

The problem of separation (not exception!) of Zitterbewegung is successfully
solved by means of quantum theory with definite parity of operators~\cite{49}.
It is found that the general classical relations like the spin mass
(\ref{m}), momentum invariant (\ref{pp}) and spin invariants (\ref{i}) have
the corresponding quantum analogues with even operators (see~\cite{28} for
$m$, see~\cite{50} for $P_\alpha P^\alpha$, and see~\cite{51} for
$H_{\mu\nu}\Pi^{\mu\nu}$).  For instance, instead of (\ref{pp}) we obtain
\begin{equation}
\stackrel{\vee}{P}_\alpha\stackrel{\vee}{P}{}^\alpha=-\stackrel{\vee}{m}c^2,
\label{wPwP}
\end{equation}
where
$$\stackrel{\vee}{m}=m_0\left(1-\frac{\mu}{2c^2}\;
H_{\alpha\beta}\stackrel{\vee}{\Pi}{}^{\alpha\beta}\right),$$
$\stackrel{\vee}{P}{}^\alpha$ and $\stackrel{\vee}{\Pi}{}^{\alpha\beta}$ are
the even operators which correspond to $\widehat{P}{}^\alpha$ and
$\sigma^{\alpha\beta}$. Operator $\stackrel{\vee}{P}{}^\alpha$ passes into
the ordinary spin operator with external field at $\hbar\to 0$, spin operator
$\stackrel{\vee}{\Pi}{}^{\alpha\beta}$ (Hilgevoord-Wouthuysen operator) does
not contain the Plank constant. It is interesting that Eq.~(\ref{wPwP}) is
invariant with respect to replacement of the degree of operators parity, i.e.
$\vee\to\vee\vee\to\vee\vee\vee\to\dots$ (see \cite{50}).

Operator equations of motion for $\stackrel{\vee}{P}{}^\alpha$ and
$\stackrel{\vee}{\Pi}{}^{\alpha\beta}$ are the Heisenberg proper time
equations with the Dirac or Dirac-Pauli squareable Hamiltonian taking into
account the anomalous magnetic moment of an electron. The obtained in this
way operator equations coincide completely in form with corresponding
classical equations at $\hbar\to 0$, but provided the
terms with the Plank constant in the latter equations
should be also tended to zero (see also quasiclassical BMT
approximation~\cite{12} et al.).  At the same time this result means that
the problem of correlation between the classical and quantum
Zitterbewegung remains vague. It is possible that this problem will be
solved via construction of a quantum theory with the invariant parity
degree of operators, or other methods.

\end{document}